\documentclass[draftclsnofoot, onecolumn, 11pt]{IEEEtran}
\usepackage[dvips]{graphicx}
\usepackage[tbtags]{amsmath}
\usepackage{amsfonts}
\usepackage{multirow}
\usepackage{latexsym}
\usepackage{amssymb}
\usepackage{amsxtra}
\newtheorem{thm}{Theorem}
\newtheorem{lmm}{Lemma}
\newtheorem{crl}{Corollary}

\begin{document}
\title{Secrecy Capacity over Correlated Ergodic Fading Channel}
\author{Hyoungsuk~Jeon,~\IEEEmembership{Student Member,~IEEE,}
        Namshik~Kim,~\IEEEmembership{Member,~IEEE,}
        Minki~Kim,~\IEEEmembership{Student member,~IEEE,}
        Hyuckjae~Lee,~\IEEEmembership{Member,~IEEE,}
        and~Jeongseok~Ha,~\IEEEmembership{Member,~IEEE}% <-this % stops a space
\thanks{Authors are with the School of Engineering, Information and
Communications University, Daejeon, Korea e-mail: \{hschun, nskim, mankigud, hjlee, jsha\}@icu.ac.kr.}}% <-this % stops a space
%-------------------------------------------------------------------%
\maketitle
\begin{abstract}
We investigate the secrecy capacity of an ergodic fading wiretap channel in which the main channel is correlated with the
eavesdropper channel. In this study, the full Channel State Information (CSI) is assumed, and thus the transmitter knows the channel gains of the legitimate receiver and the eavesdropper. By analyzing the resulting secrecy capacity we quantify the loss of the secrecy capacity due to the correlation. In addition, we study the asymptotic behavior of the secrecy capacity as Signal-to-Noise Ratio (SNR) tends to infinity. The capacity of an ordinary fading channel logarithmically increases with SNR. On the contrary, the secrecy capacity converges into a limit which can be an upper bound on the secrecy capacity over the fading wiretap channel. We find a closed form of the upper bound for the correlated Rayleigh wiretap channel which also includes the independent case as a special one. Our work shows that the upper bound is determined by only two channel parameters; the correlation coefficient and the ratio of the main to the eavesdropper channel gains that will be called PCC and CGR respectively. The analysis of the upper bound tells how the two channel parameters affect the secrecy capacity and leads to the conclusion that the excessively large signal power does not provide any advantage in the secrecy capacity, and the loss due to the correlation is especially serious in low CGR regime.
\end{abstract}
%-------------------------------------------------------------------%
%\begin{IEEEkeywords}
%Channel gain ratio, correlated channel, ergodic fading wiretap channel, power correlation coefficient, secrecy capacity, upper bound.
%\end{IEEEkeywords}
%-------------------------------------------------------------------%
\ifCLASSOPTIONpeerreview
\begin{center} \bfseries EDICS Category: 3-BBND \end{center}
\fi
%-------------------------------------------------------------------%
\IEEEpeerreviewmaketitle
\section{Introduction} \label{Sec:Intro}
\IEEEPARstart{T}{h}e notion of information-theoretic secrecy \cite{Shannon49Com} was first introduced by Shannon where he showed that the transmitter and the legitimate receiver need to share a random key of length $k$ to secure $k$ bit information from the eavesdropper. That is, the transmitted message $W$ is independent of the eavesdropper's observation $Z$; $I(W; Z) = 0$ which is called \emph{perfect secrecy}. Although the perfect secrecy provides unconditional secrecy, such a system called a \emph{one-time pad} requires a new random key for each new message. Thus, it may not be considered as a feasible solution in some practical situations.

Due to the difficulty of the secret key distribution, the secrecy issues have been usually addressed with cryptographic protocols such as the Rivest-Shamir-Adelman (RSA) scheme and Advanced Encryption Standard (AES) which instead provide \emph{computational security}. That is, to break the secrecy measures in time, the required complexity of the eavesdropper becomes prohibited with current technology. Although the concept of computational security is relatively weak as compared to the perfect secrecy, it has been widely adopted in practical systems and implemented on the application layer independent of the physical layer design.

In the meantime, Wyner also considered the information-theoretic secrecy on a channel model called \emph{wiretap channel}
\cite{Wyner75TheWiretap} where a legitimate receiver communicates over a main channel, and observations at a wiretapper \footnote{We will use the wiretapper and the eavesdropper interchangeably.} are degraded from the ones at the legitimate receiver. He showed that the information rate to the legitimate receiver and the ignorance at the wiretapper can be traded off when the wiretapper has a degraded channel. In his work, the maximum information rate of the main channel with the total ignorance at the wiretapper is defined as \emph{secrecy capacity}, and he proved the existence of channel codes achieving the secrecy capacity. Hence the perfect secrecy is now achievable without sharing random keys. After his work, there have been numerous related works \cite{Csiszar78Broadcast, Leung-Yan-Cheong77OnASpecial, Leung-Yan-Cheong78TheGaussian, Ozarow84Wiretap} for variations of the wiretap channel.

Recently, the proliferation of wireless devices has contributed to the improvement of living standards, but on the other hands caused a growing uneasiness about the leakage of private information. This insecurity may be attributed to the broadcast nature of radio propagation and the inherent randomness of wireless channel which make the radio transmission vulnerable to attacks from unexpected eavesdroppers. Thus, it seems to be a matter of course to apply the results of the wiretap channels to the secrecy of wireless communications. However, due to the nature of wireless communications, it is not always guaranteed that the eavesdropper channel is noisier than the main channel. In many cases, the eavesdropper can have even a better channel which results in zero secrecy capacity.

Soon, it is realized that the inherent randomness of wireless channels gives an opportunity to achieve a positive secrecy capacity even if the eavesdropper channel is better in the average sense. On slow fading channels, the secrecy capacity is investigated in terms of outage probability \cite{Barros06Secrecy, Bloch08Wireless}. Further studies on the secrecy capacity of wireless channels have been done in many difference aspects; the ergodic secrecy capacity of fading in \cite{Liang08Secure, Li06Secrecy, Gopala08OnTheSecrecy}, secure broadcasting in \cite{Khisti06Secure}, \cite{Dijk97OnASpecial} space-time signal processing \cite{Hero03Secure}, \cite{Shafiee07TowardsTheSecrecy}, \cite{Oggier08Secrecy} and etc.

In this correspondence we investigate the secrecy capacity of an ergodic fading wiretap channel in which the main channel is correlated with the eavesdropper channel. The ergodic fading wiretap channel was already studied in \cite{Gopala08OnTheSecrecy} where messages are transmitted opportunistically when the main channel has a better instantaneous channel gain than that of the eavesdropper channel. Thus even in the case that the main channel is noisier, due to the opportunistic transmission, a positive secrecy capacity is still achievable. However, if the two channels are correlated, such an opportunistic scheme loses the chance to transmit and thus leads to a loss of the secrecy capacity. In real radio environments, correlation between two channels is frequently observed \cite{Lee73Effects, Rhee74Results}. The level of the correlation highly depends on antenna deployments, proximity of the legitimate receiver and eavesdropper, and scatterers around them \cite{Lee73Effects, Rhee74Results, Shiu00Fading, Tse05Fund}. For example, antenna deployments at high altitude in rural or suburban area generate dominant line-of-sight paths, which results in high correlation between the two receivers. Moreover, it is also possible that the eavesdropper actively induces the correlation, e.g., by approaching the legitimate receiver. Although the correlation is a crucial channel parameter affecting the secrecy capacity, to the best of our knowledge, no previous study has been done on this topic.

Motivated by the practical scenario, we first derive the secrecy capacity for the correlated wiretap channel and analyze the impact of the correlation on the secrecy capacity, which quantitatively show how much of the secrecy capacity will be lost due to the correlation. However, we are more interested in the analytic study on the secrecy capacity with different values of channel parameters. To do so, we investigate the asymptotic behaviors of the secrecy capacity as Signal-to-Noise Ratio (SNR) tends to infinity. The capacity of an ordinary fading channel logarithmically increases
with SNR \cite{Tse05Fund}. On the contrary, the secrecy capacity converges into a limit which can be an upper bound on the secrecy capacity over the fading wiretap channel. We find a closed form of the upper bound for the correlated Rayleigh wiretap channel which also includes the independent case \cite{Gopala08OnTheSecrecy} as a special one. Our work shows that the upper bound is determined by only two channel parameters; the correlation coefficient and the ratio of the main to the eavesdropper channel gains. The analysis of the upper bound tells how the two channel parameters affect the secrecy capacity and leads to the conclusion that the excessively large signal power does not provide any advantage in the secrecy capacity. In addition, we will show that the loss due to the correlation is especially detrimental where the channel gain ratio is small. We believe our work makes the results in \cite{Gopala08OnTheSecrecy} more comprehensive and also provides a way to evaluate the required rate margin due to the active eavesdropper who intensionally induces the correlation. Although we focus on the secrecy capacity of the correlated ergodic fading wiretap channel, the analysis can be easily applicable to other scenarios such as the outage probability analysis on the slow fading channel \cite{Barros06Secrecy, Bloch08Wireless}.

The remainder of this correspondence is organized as follows. In Section \ref{Sec:System-Model}, we describe the system model considered in our work. The secrecy capacity for the correlated ergodic fading channel is presented in Section \ref{Sec:Secrecy-Cap}. The upper bound of the formulated secrecy capacity is also derived in a closed-form expression in Section \ref{Sec:Secrecy-Cap}. In Section \ref{Sec:Num-Results} we present the numerical results and discuss the relation between the correlation and the loss of the secrecy capacity. Finally, we summarize our results in Section \ref{Sec:Conclusion}.

%-------------------------------------------------------------------%
\section{System model} \label{Sec:System-Model}
\begin{figure}[!t]
\center
\includegraphics[width=5.5in]{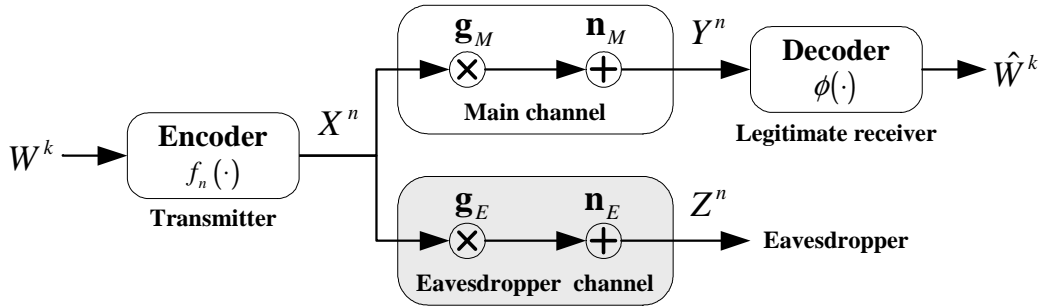}\\
\caption{System model} \label{fig:fig1}
\end{figure}
Let us consider a fading wiretap channel model depicted in Fig. \ref{fig:fig1}. A transmitter constructs an $\left(M, n\right)$ code and wishes to send the message to a legitimate receiver with an arbitrarily low probability of error, while securing against eavesdropping of an unintended user. Specifically, the transmitter maps confidential messages $W \in \mathcal{W}=\{1,\ldots,M\}$ to a codeword $x^n \in \mathcal{X}^n$ by using a stochastic encoder $f_n \left(\cdot\right):\mathcal{W}\rightarrow\mathcal{X}^n$. Then, the received signals of the legitimate receiver and the eavesdropper at the $i$-th coherent time are given as follows:
\begin{align*}
  y\left(i\right)& = g_{M}\left(i\right)x\left(i\right) + n_{M}\left(i\right)\\
  z\left(i\right)& = g_{E}\left(i\right)x\left(i\right) + n_{E}\left(i\right),
\end{align*}
where $n_{M}\left(i\right)$ and $n_{E}\left(i\right)$ are the independent and identically distributed (i.i.d.) Gaussian noise with zero mean and unit variance, and $g_{M}\left(i\right)$ and $g_{E}\left(i\right)$ denote the channel gains of the main and eavesdropper channels respectively. We assume that the main channel is correlated with the eavesdropper channel, and the both are ergodic block fading channels. The legitimate receiver then decodes received signals $y^n \in \mathcal{Y}^n$ by using a function $\phi \left(\cdot\right):\mathcal{Y}^n\rightarrow\mathcal{W}$. Let $\hat w = \phi \left(y^n\right)$ be the estimated messages at the legitimate receiver, then the average error probability of an $\left(M,
n\right)$ code is defined as
\[
  P_e^n  = \frac{1}{M}\sum\limits_{w \in \mathcal{W}} \Pr \left( \hat w  \ne w |w \text{ is sent} \right).
\]

Let us denote the power gains of the main and eavesdropper channels as $h_M \left(i\right)= \left|g_M \left(i\right) \right|^2$ and $h_E \left(i\right)= \left|g_E \left(i\right)\right|^2$ respectively and assume the full channel state information (CSI) at the transmitter. Then the equivocation rate which measures the secrecy level of confidential messages against the eavesdropper is defined as
\[
  R_e \triangleq \frac{1}{n} H\left(W |Z^n , h_M^n ,h_E^n\right),
\]
where $h_M^n$ and $h_E^n$ are the vectors of the power gains for the main and the eavesdropper channels. Adopting the definition from \cite{Gopala08OnTheSecrecy, maurer00informationtheoretic, Barros08Strong}, we say that the rate $R_s$ is achievable with weak secrecy if, for any given $\epsilon > 0$, there exists a $\left(2^{nR_s},n\right)$ code of sufficient large $n$ such that
\begin{align*}
  P_e^n  & \le  \epsilon  \\
  R_e    & \ge  R_s  - \epsilon.
\end{align*}
The secrecy capacity is then the supremum of achievable secret rates
\[
  C_s\triangleq\mathop {\sup }\limits_{P_e^n  \le \epsilon } \left\{R_s: R_s \text{ is achievable}\right\}.
\]

%-------------------------------------------------------------------%
\section{Secrecy capacity over correlated channels} \label{Sec:Secrecy-Cap}
We begin with introducing the secrecy capacity when the main channel is correlated with the eavesdropper channel. Let $f_{H_M, H_E}\left( {h_M ,h_E } \right)$ be the joint probability density function (pdf) of $H_M$ and $H_E$, which are random variables of the fading power gains for the main and the eavesdropper channels respectively. Assuming that the perfect CSIs of both channels are available at the transmitter, we modified the theorem in \cite{Gopala08OnTheSecrecy} as follows.
%-------------------------------------------------------------------%
%                             Theorem 1                             %
%-------------------------------------------------------------------%
\begin{thm}[Gopala'08]
When the main and the eavesdropper channels are correlated with each other, the secrecy capacity is given by
\begin{align}\label{eq:thrm1}
  C_s  = \max\limits_{P\left( {h_M ,h_E } \right)} \int_0^\infty  \int_{h_E }^\infty  & \big[ \log \left( 1 + h_M P\left( {h_M ,h_E } \right) \right) - \log \left( 1 + h_E P\left( h_M ,h_E \right) \right) \big] \\\nonumber
  & \times f_{H_M, H_E}\left(h_M, h_E\right)dh_M dh_E \\\nonumber
  & \text{such that } \mathbb{E}\left\{ {P\left( {H_M ,H_E } \right)} \right\} \le \bar P
\end{align}
\end{thm}
%-------------------------------------------------------------------%
%                         Theorem 1 - Proof                         %
%-------------------------------------------------------------------%
\begin{IEEEproof}
We follow the proof in \cite{Gopala08OnTheSecrecy} and describe only the places to be modified to include the correlation. The main idea in \cite{Gopala08OnTheSecrecy} is the opportunistic transmission with a rate adaptation over the quantized fading channel. Specifically, it first quantizes the main and the eavesdropper channel gains into finite bins and then regards a quantized channel state as a time-invariant additive white Gaussian noise (AWGN) wiretap channel. Thus the existence of a coding scheme to achieve the secrecy capacity at any instant is guaranteed by the coding theorem of the AWGN wiretap channel in \cite{Leung-Yan-Cheong78TheGaussian}. Averaging over all channel states, the achievability of the secrecy capacity for the ergodic fading channel is finally proved. This average secrecy rate is computed by as follows:
\[
  R_s  = \sum\limits_i \sum\limits_j \left(R_s \right)_{ij} \Pr \left(H_M  \approx h_{M,i}, H_E  \approx h_{E, j}\right),
\]
where $h_{M,i}$ and $h_{E,j}$ are the $i$-th and the $j$-th quantized channel states of the main and the eavesdropper channel respectively, and $\left(R_s\right)_{ij}$ is the secrecy rate of a time-invariant AWGN wiretap channel with channel gains $h_{M,i}$ and $h_{E,j}$. For the correlation scenario, $\Pr \left( {H_M  \approx h_{M,i} , H_E  \approx h_{E,j} } \right)$ is not from the product of marginal pdfs of $H_{M}$ and $H_{E}$ but from the joint pdf of $H_{M}$ and $H_{E}$. The remaining part of the proof in \cite{Gopala08OnTheSecrecy} is the same for the correlation scenario.
\end{IEEEproof}
%-------------------------------------------------------------------%
\subsection{Upper bounds of the secrecy capacity}
It is well known that the capacity of the wireless channel without secrecy constraints highly depends on the received power. If other resources such as the bandwidth and channel gains are fixed, the capacity logarithmically increases with the SNR. In other words, the capacity has been on the increase with the SNR, even though the effect of the SNR on the capacity gets smaller due to the concavity of a logarithm function. However, for the wiretap channel the secrecy capacity converges into a certain value. This behavior is in a striking contrast with the capacities of ordinary communication channels and thus the excessively large signal power does not affect the secrecy capacity at all. Natural questions are then what is the limiting value to which the secrecy capacity eventually converges as the SNR increases and how this limit depends on other resources and/or parameters in the wiretap channel.

Let us rewrite the secrecy capacity of correlated channels from
(\ref{eq:thrm1}) as
\begin{equation*}\label{eq:thrm1_ver2}
  C_s = \max \limits_{P\left(H_M, H_E \right)} \mathbb{E}_{H_M > H_E} \left[ \log \left( 1 + H_M P \left( H_M, H_E  \right) \right) - \log \left( 1 + H_E P\left( H_M, H_E \right) \right) \right],
\end{equation*}
where in general, the secrecy capacity is a result of the power allocation strategy, $\max_{P(h_M, h_E)}$. However, at high SNR regime, using $\log\left(1 + x\right) \approx \log(x)$ for large $x$, we get the secrecy capacity as
follows:
\begin{equation}\label{eq:apprx1}
  C_s \approx \mathbb{E}_{H_M  > H_E } \left[ {\log \left( {\frac{{H_M}} {{H_E }}} \right)} \right]  = C^{\lim}_s,
\end{equation}
which clearly shows that as the signal power grows the secrecy capacity is determined by the channel gain ratio regardless of the power allocation strategy. We regard the limit of secrecy capacity in (\ref{eq:apprx1}) as an upper bound \footnote{$C_s$ is monotonically increasing function with respect to $P\left(H_M,H_E\right)$ since its derivative is always positive where $H_M > H_E$. Thus $C_s$ is bounded by $C_s^{\lim} \triangleq \lim_{P\left(H_M,H_E\right) \to \infty } C_s$.} of the secrecy capacity and study the behaviors of the upper bound with different values of the channel gain ratio and channel correlation.

Under the Rayleigh fading assumption, we now derive the limiting value (upper bound) of the secrecy capacity for the wiretap channel in a closed form. To this end, let $U = {{H_M }}/{{H_E }}$. We will prove in the following lemma that the pdf of $U$ is determined by the average Channel power Gain
Ratio (CGR), $\kappa=\mathbb{E}[H_{M}]/\mathbb{E}[H_{E}]$, and the Power Correlation Coefficient (PCC), $\rho$ between $H_{M}$ and $H_{E}$. Then the upper bound of the  secrecy capacity, i.e.,
$C_s^{\lim}\left(\kappa,\rho\right)$, in (\ref{eq:apprx1}) can be expressed in terms of a single random variable $U$:
\begin{equation}\label{eq:apprx1_ver2}
  C_s^{\lim}\left(\kappa,\rho\right) = \int_1^\infty  {\log u\,f_U \left( u \right)du}.
\end{equation}
To solve (\ref{eq:apprx1_ver2}), we first introduce the following lemma.
%-------------------------------------------------------------------%
%                              Lemma 2                              %
%-------------------------------------------------------------------%
\begin{lmm} \label{lmm:pdfU}
  Let $H_{M}$ and $H_{E}$ be the correlated exponential distributions. Then $U=H_{M}/H_{E}$ has the pdf given as
\begin{equation*}\label{eq:pdfU}
  f_U \left( u \right)= \kappa \frac{{\left( {1 - \rho } \right)\left({u + \kappa } \right)}}{{\left[ {\left( {u + \kappa } \right)^2  - 4 \rho \kappa u } \right]^{3/2} }}.
\end{equation*}
\end{lmm}
%-------------------------------------------------------------------%
%                          Lemma 2 - Proof                          %
%-------------------------------------------------------------------%
\begin{IEEEproof}
Suppose $H_E = h_E$ is given, then $U = {{H_M }}/{{h_E }}$ is a function of a single random variable $H_M$. Therefore the conditional pdf of $U$ is $f_U \left(u|h_E\right)=h_E f_{H_M}\left(uh_E |h_E\right)$ for $h_E \ge 0$. Then the pdf of $U$ is expressed in terms of the conditional pdf as
\begin{align}
  \label{eq:pdf_joint}\nonumber
  f_U \left( u \right) & = \int {f_U \left( {u|h'_E } \right)f_{H_E } \left( {h'_E } \right)dh'_E}\\\nonumber
   & = \int {{h'_E } f_{H_M } \left( {uh'_E |h'_E } \right)f_{H_E } \left( {h'_E } \right)dh'_E}\\
   & = \int {h'_E f_{H_M H_E } \left( {uh'_E ,h'_E }\right)dh'_E}.
\end{align}

Let $\bar \gamma _M=\mathbb{E}[H_{M}]$ and $\bar \gamma
_E=\mathbb{E}[H_{E}]$. By applying the joint pdf of $H_M$ and $H_E$
in Appendix to (\ref{eq:pdf_joint}), we have
\begin{align*}
f_U \left( u \right) & = \int_0^\infty {\frac{h'_E}{{\bar \gamma _M \bar \gamma _E \left( {1 - \rho }\right)}} \exp \left[ { - \frac{h'_E}{{ {1 - \rho } }}\left({\frac{u}{{\bar \gamma _M }} + \frac{1}{{\bar \gamma _E }}} \right)} \right]I_0 \left( { \frac{{2 h'_E}}{{{1 - \rho } }}\sqrt {\frac{\rho u}{{\bar \gamma _M \bar \gamma _E }}}} \right)dh'_E }\\
  & \mathop  = \limits^{\left( a \right)}  \frac{1}{{\bar \gamma _M \bar \gamma _E \left( {1 - \rho } \right)}}\frac{\frac{1}{{{1 - \rho }}}\left( {\frac{u}{{\bar \gamma _M }} + \frac{1}{{\bar \gamma _E }}} \right) }{{\left[ {\left(\frac{1}{{{1 - \rho  }}}\left( {\frac{u}{{\bar \gamma _M }} + \frac{1}{{\bar \gamma _E }}} \right) \right)^2  - \left(\frac{{2}}{{{1 - \rho }}}\sqrt {\frac{\rho u}{{\bar \gamma _M \bar \gamma _E }}}\right)^2 } \right]^{3/2} }}\\
  & = \frac{{\bar \gamma _M }}{{\bar \gamma _E }}\frac{{\left( {1 - \rho } \right)\left( {u + \frac{{\bar \gamma _M }}{{\bar \gamma _E }}} \right)}}{{\left[ {\left( {u + \frac{{\bar \gamma _M }}{{\bar \gamma _E }}} \right)^2  - 4\rho\frac{{\bar \gamma _M }}{{\bar \gamma _E }}  }u \right]^{3/2} }}
  = \kappa \frac{{\left( {1 - \rho } \right)\left({u + \kappa } \right)}}{{\left[ {\left( {u + \kappa } \right)^2  - 4 \rho \kappa u } \right]^{3/2} }},
\end{align*}
where $I_0 \left( x \right) \buildrel \Delta \over = \frac{1}{{2\pi}}\int_0^{2\pi } {e^{x\cos \theta } d\theta }$ is the zero order modified bessel function of the first kind and $\left(a\right)$ follows from the table of integrals in
\cite{Gradshteyn00TableOfIntegrals},
\[
  \int_0^\infty {x \exp \left( { - \alpha x} \right)I_0 \left( {\beta x} \right)dx}  = \frac{\alpha}{{\left( {\alpha ^2  + \beta ^2 } \right)^{\frac{3}{2}} }},
\]
for $\rm{Re}\{\alpha\} > |\rm{Re} \{\beta\}|$. Finally, we replace $\bar \gamma_M/ \bar \gamma_E$ with $\kappa$ and finish the proof.
\end{IEEEproof}
%-------------------------------------------------------------------%

Since we have the pdf of $U$ in Lemma \ref{lmm:pdfU}, we can compute the secrecy capacity $C^{\lim}_s(\kappa, \rho)$ in (\ref{eq:apprx1_ver2}). After series of mathematical manipulations, we find the limit of secrecy capacity in a closed form which is summarized in Theorem \ref{thm:apprx1_final}. The resulting secrecy capacity in (\ref{eq:apprx1_final}) consists of two terms, and we can clearly see how CGR and PCC contribute to the secrecy capacity. The first term is the limit of secrecy capacity when the channels are independent and thus depends only on CGR. On the other hand, the second term explains the loss due to PCC. In Section \ref{Subsec:Asym} we will discuss details of the results in Theorem \ref{thm:apprx1_final} and have more insights into how the two channel parameters contribute to the secrecy capacity

%-------------------------------------------------------------------%
%                             Theorem 2                             %
%-------------------------------------------------------------------%
\begin{thm} \label{thm:apprx1_final}
If the main channel is correlated with the eavesdropper channel, and the joint pdf of them is bivariate Rayleigh distribution, as SNR increases, the secrecy capacity converges into the following limiting value
\begin{equation}\label{eq:apprx1_final}
  C_s^{\lim} \left(\kappa,\rho\right)= \log \left( {1 + \kappa } \right) + \log \left( {\frac{1}{2} + \sqrt {\frac{1}{4} - \frac{{\rho \kappa }}{{\left( {1 + \kappa } \right)^2 }}} } \right).
\end{equation}
\end{thm}
%-------------------------------------------------------------------%
%                         Theorem 2 - Proof                         %
%-------------------------------------------------------------------%
\begin{IEEEproof}
It is possible to express the upper bound of the secrecy capacity $C_s^{\lim}(\kappa, \rho)$ in (\ref{eq:apprx1_ver2}) as
\begin{equation}
  C_s^{\lim} \left(\kappa,\rho\right) =  \left[ {\log u F_U\left( u \right)} \right]_1^\infty - \int_1^\infty  {\frac{1}{u} F_U\left( u \right)du}.\label{eq:part_int}
\end{equation}
with the integration by parts rule where $F_U\left( u \right)$ is the indefinite integral of $f_U\left(u\right)$. From Lemma \ref{lmm:pdfU}, $F_U\left( u \right)$ is given by
\begin{align}\nonumber
  F_U \left( u \right) & = \int {f_U\left( u \right)du}\\[2mm]
                       & = \frac{{u - \kappa }}{{2\sqrt {\left( {u + \kappa } \right)^2 - 4 \rho \kappa u} }}.\label{eq:cdfU}
\end{align}
The indefinite integral of the second term on the right side of (\ref{eq:part_int}) is
\begin{align}
  \nonumber
  \int  {\frac{1}{u} F_U\left( u \right)du} & = \int {\frac{1}{u}\frac{{u - \kappa }}{{2\sqrt {\left( {u + \kappa }
\right)^2  - 4 \rho \kappa u} }}du}\\[2mm]
  & =  \frac{1}{2}\log \left( { - \frac{{\Phi\left( u \right)}}{{\kappa u}}} \right),\label{eq:second}
\end{align}
where ${\Phi\left( u \right)}={4\left( {1-\rho } \right)\left(
{\left( {u + \kappa } \right)^2  + \left( {u + \kappa } \right)\sqrt
{\left( {u + \kappa } \right)^2  - 4 \rho \kappa u } - 2 \rho \kappa
u } \right)}$. By substituting (\ref{eq:cdfU}) and
(\ref{eq:second}) into (\ref{eq:part_int}), we finally have
the secrecy capacity limit as follows.
\begin{align*}
  C_s^{\lim} \left(\kappa,\rho\right) & = \left[ \frac{{ \left(u - \kappa \right)\log u}}{{2\sqrt {\left( {u + \kappa } \right)^2 - 4 \rho \kappa u} }} \right]_{1}^{\infty }  - \left[ {\frac{1}{2}\log \left( { - \frac{{\Phi \left( u \right)}}{{\kappa u}}} \right)} \right]_{1}^{\infty }\\
  & =  \left[ {\frac{{\left(1 - \frac{\kappa }{u}\right)\log u}}{{2\sqrt {1 + \frac{{2\kappa \left( {1 - 2\rho} \right)}}{u} + \frac{{\kappa ^2 }}{{u^2 }}} }} - \frac{1}{2}\log \left( { - \frac{{\Phi \left( u \right)}}{{\kappa u}}} \right)} \right]_{u=\infty } + \left[ {\frac{1}{2}\log \left( { - \frac{{\Phi \left( u \right)}}{{\kappa
u}}} \right)} \right]_{u=1} \\[2mm]
  & = \left[ {\frac{1}{2}\log \left( { - \frac{{\kappa }}{{\Phi \left( u \right)/u^2}}} \right)} \right]_{u=\infty }  + \left[ {\frac{1}{2}\log \left( { - \frac{{\Phi \left( u \right)}}{{\kappa
u}}} \right)} \right]_{u=1}\\[2mm]
  & = \frac{1}{2}\log \left[ {\frac{{\left( {1 + \kappa} \right)^2 + \left( {1 + \kappa } \right)\sqrt {\left( {1 + \kappa } \right)^2 - 4\rho\kappa  }  - 2 \rho\kappa }}{2}} \right]\\
  & = \log \left( {1 + \kappa } \right) + \log \left( {\frac{1}{2} + \sqrt {\frac{1}{4} - \frac{{\rho \kappa }}{{\left( {1 + \kappa } \right)^2 }}} } \right).% \label{eq:apprx1_ver3}.
\end{align*}
\end{IEEEproof}
%-------------------------------------------------------------------%
\subsection{Asymptotic properties of the secrecy capacity} \label{Subsec:Asym}
Our primary interest is then the impact of CGR on the limiting value of the secrecy capacity. To examine the asymptotic behavior of the secrecy capacity, we first consider two extreme cases: 1) independent case $\rho = 0$ and 2) completely correlated case $\rho = 1$. When the main and eavesdropper channels are independent, i.e., $\rho = 0$, $C_s^{\lim}$ in (\ref{eq:apprx1_final}) becomes
\begin{align}\label{eq:up_bound}
  C_s^{\lim} \left(\kappa, 0\right) & = \log\left(1 + \kappa\right),\\
  & \approx
  \begin{cases} \label{eq:up_bound2}
    \kappa  & \text{ for } \kappa \ll 1,\\
    \log \kappa      & \text{ for } \kappa \gg 1
  \end{cases}
\end{align}
which depends only on CGR. This is analogous to the capacity formula of an AWGN channel without secrecy constraints if CGR is regarded as the SNR in the capacity formula for the AWGN channel. In (\ref{eq:up_bound2}), we approximate $C^{\lim}_s$ at high and low CGR regimes and find that $C^{\lim}_s$ is linearly proportional to CGR at low CGR regime but it becomes logarithmic at high CGR regime.

The result in Theorem \ref{thm:apprx1_final} also shows that the effect of the correlation is only detrimental. Note that the second term in (\ref{eq:apprx1_final}) represents the loss due to the correlation and it is in the range from $-\log \frac{1}{2}$ to $0$ since $0 \le \rho < 1$ and $\kappa \ge 0$. In the worst case that PCC approaches one, i.e., the completely correlated scenario, the limiting value of the secrecy capacity in (\ref{eq:apprx1_final}) is given by
\begin{equation}
  \lim_{\rho \to 1 } C_s^{\lim} \left(\kappa, \rho\right) =
  \begin{cases}
    \log \kappa , & \text{for $\kappa > 1$}\\
               0, & \text{for $0 \le \kappa  \le 1$}.
  \end{cases} \label{eq:lw_bound}
\end{equation}
Worthy of note is that there still exists the positive secrecy capacity when $\kappa > 1$ even if the channels are completely correlated. This result can be interpreted by considering the AWGN wiretap channel in \cite{Leung-Yan-Cheong78TheGaussian}. Although the statistics of both channels are identical ($\rho \rightarrow 1$), the power gain of the main channel is larger than that of eavesdropper channel ($\kappa > 1$). Thus it can be viewed as the AWGN wiretap channel where the received SNR of the legitimate receiver is larger than that of the eavesdropper. This always provides the transmitter with the opportunity to send the secret messages, which explains how the positive secrecy is achievable even in the completely correlated case. For $0 < \rho <1$, the limiting value of the secrecy capacity is bounded by (\ref{eq:up_bound}) and (\ref{eq:lw_bound}).

To investigate the relative loss with respect to the independent case where the secrecy capacity limit is maximized, we find upper and lower bounds of the secrecy capacity limit $C^{\lim}_s(\kappa, \rho)$ in terms of $C_s^{\lim}(\kappa, 0)$ and $\rho$. Such bounds are summarized in Corollary \ref{crl:upper-and-lower-bound}.
%-------------------------------------------------------------------%
%                            Corollary 1                            %
%-------------------------------------------------------------------%
\begin{crl} \label{crl:upper-and-lower-bound}
For given $\kappa$ and $\rho$, $C_s^{\lim} \left(\kappa,\rho\right)$
is bounded by
\[
  \left(1-\rho\right)C_s^{\lim} \left(\kappa,0\right) \le C_s^{\lim} \left(\kappa,\rho\right) \le C_s^{\lim} \left(\kappa,0\right).
\]
\end{crl}
%-------------------------------------------------------------------%
%                        Corollary 1 - Proof                        %
%-------------------------------------------------------------------%
\begin{IEEEproof}
The proof is equivalent to showing that
\begin{eqnarray*}
1-\rho \le \frac{C_s^{\lim} \left(\kappa,\rho\right)}{C_s^{\lim}
\left(\kappa,0\right)} \le 1.
\end{eqnarray*}
It is easily verified that $C_s^{\lim} \left(\kappa,\rho\right)/C_s^{\lim} \left(\kappa,0\right)$ monotonically increases with CGR ($\kappa \ge 0$). Thus we can obtain the lower and upper bounds by letting $\kappa \rightarrow 0$ and $\kappa \rightarrow \infty$ respectively. First, we can see that $C_s^{\lim}(\kappa, \rho)$ in (\ref{eq:apprx1_final}) tends to $\log(1 + \kappa) = C_s^{\lim} \left(\kappa,0\right)$ as $\kappa$ increases, which gives us the upper bound. By applying the L'H\^{o}pital's rule to $\lim_{\kappa \rightarrow 0} C_s^{\lim}(\kappa, \rho)/C_s^{\lim}(\kappa, 0)$, we have the lower bound as
\[
  \lim_{\kappa \to 0} \frac{\log(1 + \kappa) + \log\left(\frac{1}{2} + \sqrt{\frac{1}{4} - \frac{\rho \kappa}{(1 + \kappa)^2}}\right)}{\log(1 + \kappa)} = 1 - \rho
\]
which finishes this proof.
\end{IEEEproof}
%-------------------------------------------------------------------%

  In this section we have seen the asymptotic behaviors of the secrecy capacity in a few limiting situations. Although such analysis gives insights into how the channel parameters affect the secrecy capacity we are also interested in the secrecy capacity at moderate SNR values. In the next section, we will evaluate the secrecy capacity in a quantitative manner.

%-------------------------------------------------------------------%
\section{Numerical results} \label{Sec:Num-Results}
\begin{figure}[!t]
\centering
\includegraphics[width=5.5in]{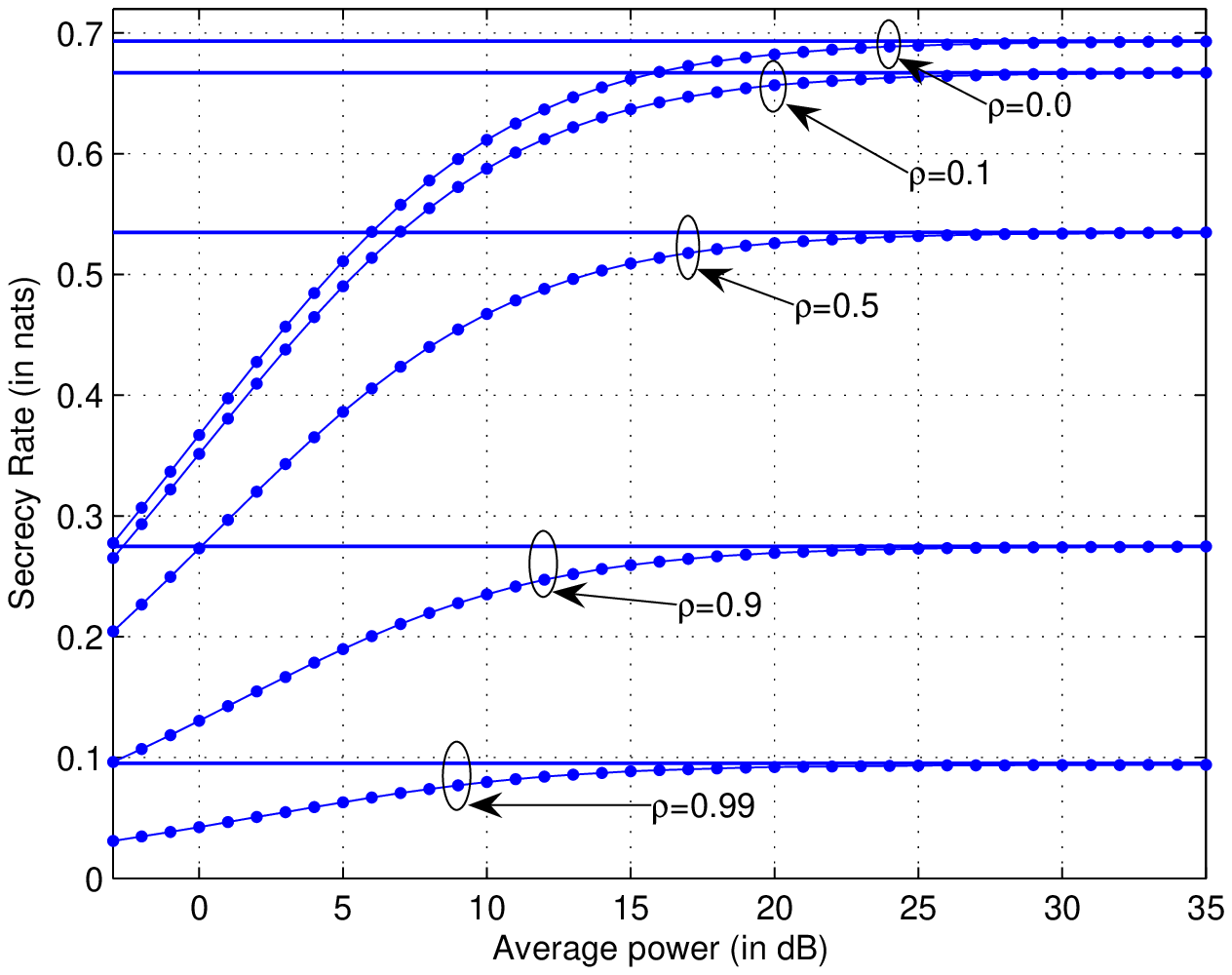}\\
\caption{The perfect secrecy rate as a function of average power for the symmetric case (CGR=1.0); The solid lines indicate the upper bounds on secrecy capacity in (\ref{eq:apprx1_final}), and the lines with filled circles represent the numerical evaluations of the secrecy capacity in (\ref{eq:thrm1}).}
\label{fig:fig2}
\end{figure}

\begin{figure}[!t]
\centering
\includegraphics[width=5.5in]{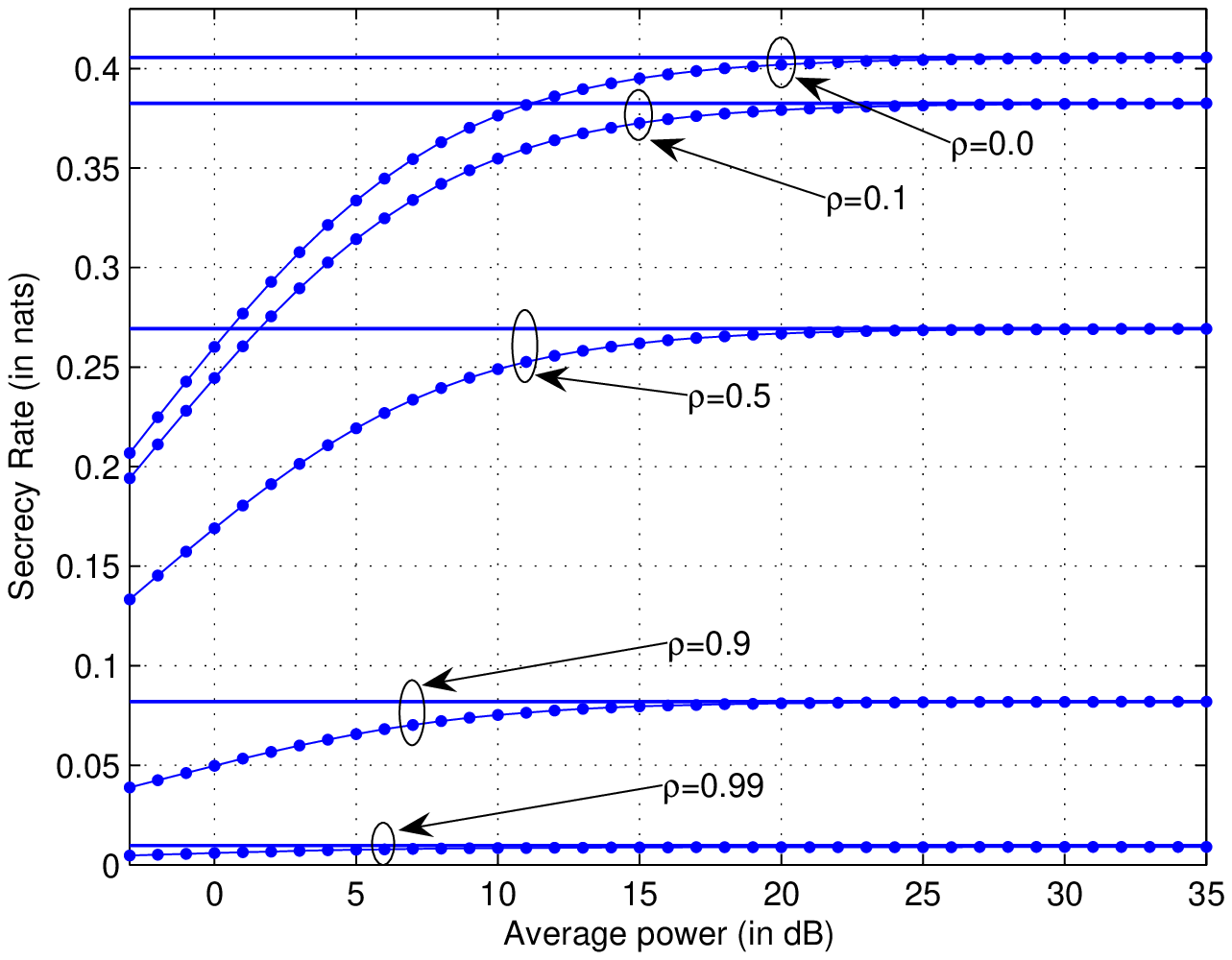}\\
\caption{The perfect secrecy rate as a function of average power for the asymmetric case (CGR=0.5); The solid lines indicate the upper bounds on secrecy capacity in (\ref{eq:apprx1_final}), and the lines with filled circles represent the numerical evaluations of the secrecy capacity in (\ref{eq:thrm1}).}
\label{fig:fig3}
\end{figure}

\begin{figure}[!t]
\centering
\includegraphics[width=5.5in]{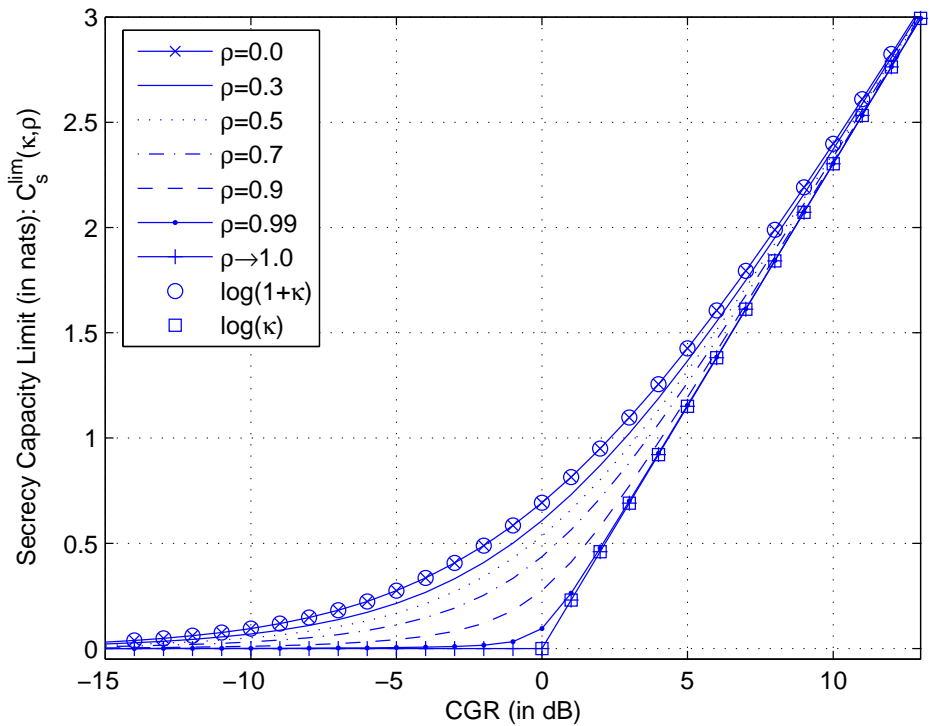}\\
\caption{The secrecy capacity limit versus CGR}
\label{fig:fig4}
\end{figure}

\begin{figure}[!t]
\centering
\includegraphics[width=5.5in]{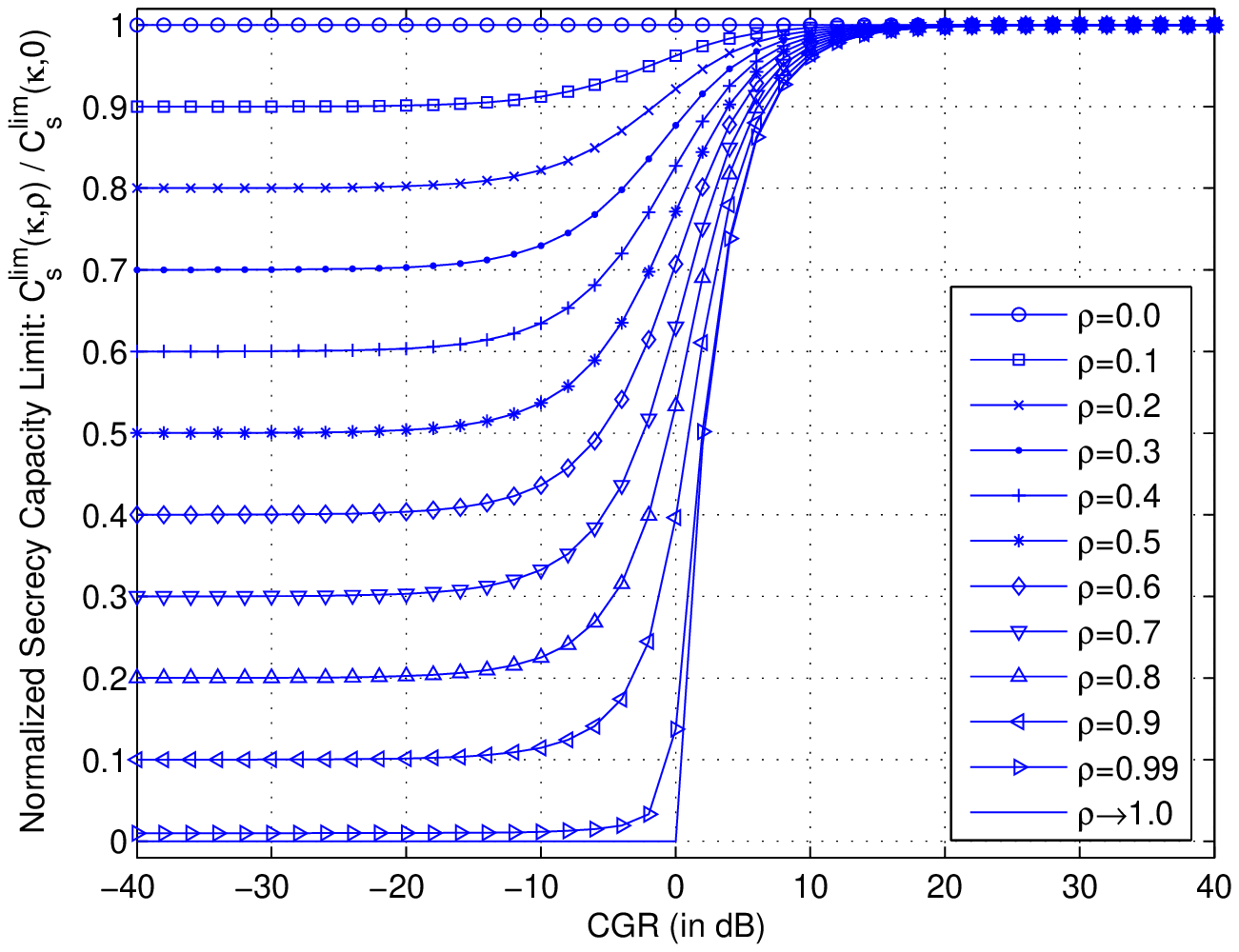}\\
\caption{Normalized secrecy capacity limit versus CGR }
\label{fig:fig5}
\end{figure}

In this section, we evaluate the secrecy capacity for different PCC values in a range of SNR values and confirm the analytic results in Section \ref{Subsec:Asym}. We also discuss how an active eavesdropper can take advantage of the correlation to decrease the secrecy capacity.

To evaluate the secrecy capacity $C_s$ in (\ref{eq:thrm1}), the instantaneous signal power $P(h_M, h_E)$ must be determined in a way to maximize the information rate for the average power constraint,
\[
  \mathbb{E}\left\{ {P\left( {H_M ,H_E } \right)} \right\} \le \bar P.
\]
In such a way, we use the method of Lagrange multipliers and have the following optimal power allocation strategy \cite{Gopala08OnTheSecrecy}:
\begin{equation}\label{eq:opt_power}
  P\left(h_M,h_E\right)=\frac{1}{2} \left[\sqrt{\left(\frac{1}{h_E}-\frac{1}{h_M}\right)^2+\frac{4}{\lambda}\left(\frac{1}{h_E}-\frac{1}{h_M}\right)}-\left(\frac{1}{h_M}+\frac{1}{h_E}\right)\right]^{+},
\end{equation}
where $\left[x\right]^{+}=\max\left\{0,x\right\}$ and $\lambda$ is a Lagrange multiplier determined by the power constraint. Applying the power allocation $P(h_M, h_E)$ in (\ref{eq:opt_power}) to the secrecy capacity formula in (\ref{eq:thrm1}), we numerically evaluate the secrecy capacity  at each SNR value with two CGR values: 1) for the symmetric case and 2) for the asymmetric case. The evaluations for the symmetric (CGR = 1.0) and asymmetric (CGR = 0.5) cases are depicted in Figs. \ref{fig:fig2} and \ref{fig:fig3}, respectively where the curves for $\rho = 0$ correspond to the results in \cite{Gopala08OnTheSecrecy}. To confirm our work in Section \ref{Subsec:Asym}, we also evaluate the limits of the secrecy capacity in (\ref{eq:apprx1_final}) and compare them with the secrecy capacity curves. The comparisons show that the secrecy capacity curves with the different PCC values converge into the limit of the secrecy capacity $C^{\lim}_s(\kappa, \rho)$. Thus, the analytic results in Section \ref{Subsec:Asym} is confirmed by the numerical evaluations. As aforementioned, such convergence implies that the signal power becomes more inefficient as it grows, and eventually the secrecy capacity is independent of the signal power.

  Figs. \ref{fig:fig4} and \ref{fig:fig5} show the impact of CGR and PCC on the limiting value of the secrecy capacity. In Fig. \ref{fig:fig4}, the limit of the secrecy capacity $C_s^{\lim}(\kappa, \rho)$ in (\ref{eq:apprx1_final}) is evaluated at a few PCC values where $C_s^{\lim}(\kappa, \rho)$ has different behaviors in low and high CGR regimes. In low CGR regime, the correlation significantly degrades the limit of the secrecy capacity, which is predicted by lower bound in Corollary \ref{crl:upper-and-lower-bound}. Since the loss due to the correlation is especially serious in low CGR regime an active eavesdropper efficiently decreases the secrecy capacity by approaching the legitimate receiver, which thus results in not only high PCC but also low CGR. Equivalently, if the transmitter does not know or underestimate the correlation, the overestimated secrecy capacity results in information leak to the eavesdropper. Thus, the transmission rate must be determined in a conservative way to consider the possible correction.

  On the contrary, the impact of the correlation becomes negligible as CGR increases, and all curves for the correlated fading scenario eventually approach the secrecy capacity limit of i.i.d. fading scenario. Thus, the correlation is not an efficient way to break the security in high CGR regime. In Fig. \ref{fig:fig4}, it is also noticed that even if the channel are completely correlated, we have a positive secrecy capacity when CGR is larger than one (0 dB) as we mentioned in Section \ref{Subsec:Asym}.

  In Corollary \ref{crl:upper-and-lower-bound}, we show that the limit of secrecy capacity is low bounded by $(1 - \rho) C^{\lim}_s(\kappa, 0)$ in low CGR regime. To confirm this we depict the normalized loss of the secrecy capacity limit $C^{\lim}_s(\kappa, \rho)/C^{\lim}_s(\kappa, 0)$ in Fig. \ref{fig:fig5} where the normalized loss exactly  follows the bound $1 - \rho$. In addition, in high CGR regime, the normalized loss converges into one, which confirms the results in Corollary \ref{crl:upper-and-lower-bound}.

%-------------------------------------------------------------------%
\section{Conclusion} \label{Sec:Conclusion}

  We investigate the secrecy capacity of an ergodic fading wiretap channel in which the main channel is correlated with the eavesdropper channel. In this study, the full Channel State Information (CSI) is assumed, and thus the transmitter knows the channel gains of the legitimate receiver and the eavesdropper. To see the detrimental effect of the correlation, we find the joint probability density function of the correlated Rayleigh fading wiretap channel with which we evaluate the secrecy capacity. In the evaluation, it is noticed that the secrecy capacity converges into a limit with the growing signal-to-noise ratio as opposed to ever increasing capacity of conventional communication channels. Since it is also interesting to see the roles of the channel parameters in the secrecy capacity, we try to find the limit of the secrecy capacity in a closed form and extensively study the behaviors of the limit in various situations.

  Our study tells that the limit of the secrecy capacity determined by the two channel parameters; average channel gain ratio (CGR) and power correlation coefficient (PCC).  The study also shows that the correlation is especially detrimental when CGR is small. Thus, by approaching a legitimate receiver an active eavesdropper can efficiently incapacitate the wiretap channel codes even if the transmitter can afford high transmit signal power since such close proximity leads to low CGR and high PCC. To get more insight in low CGR regime, we find a lower bound of the secrecy capacity limit which has a linear relation with PCC. That is, the secrecy capacity linearly degrades with increasing correlation, and we confirm that the lower bound is tight enough especially in low CGR regime. This result implies that a margin of transmission rate for confidential messages must be taken into account to cope with possible correlation caused by an active eavesdropper. In such efforts our work provides a criterion to decide the rate margin.

  On the other hand, the correlation does not affect the secrecy capacity when CGR is high. Thus, although we do not propose a specific way here, our study indicates that the most efficient way to defeat the active eavesdropper is to improve CGR, which will be pursued in our future research.

  The analysis on the limit of the secrecy capacity is confirmed by evaluating the secrecy capacity in a numerical way and comparing them with the analytic results. Although the correlation is one of important parameters, to the best of our knowledge, the effects of the correlation on wiretap channel codes have not been investigated. We believe that our work paves the way for a new study on the correlation wiretap channel.

%-------------------------------------------------------------------%
\appendix[The joint pdf of $H_M$ and $H_E$]
Let the random variables $R_M$ and $R_E$ be the envelopes of complex Gaussian random variables, $G_M$ and $G_E$, respectively. The joint pdf of correlated random variables $R_M$ and $R_E$ is then the bivariate Rayleigh distribution which is given by \cite{Simon04Dig},
\begin{eqnarray*}
  f_{R_M, R_E} \left( r_M, r_E \right) = \frac{{4 r_M r_E }}{{\bar \gamma _M \bar \gamma _E \left( {1 - \rho } \right)}}\exp \left[ { - \frac{{1 }}{{{1 - \rho } }}\left( {\frac{r_M^2}{{\bar \gamma _M }} + \frac{r_E^2}{{\bar \gamma _E }}} \right)} \right]I_0 \left( {\frac{{2 \sqrt{ \rho} r_M r_E}}{{ {1 - \rho }\sqrt{\bar \gamma _M \bar \gamma _E}}} } \right),
\end{eqnarray*}
where $I_0 \left( x \right) \buildrel \Delta \over = \frac{1}{{2\pi }}\int_0^{2\pi } {e^{x\cos \theta } d\theta }$, $\bar \gamma_M=\mathbb{E}\left[R_M^2\right]$, $\bar \gamma_E=\mathbb{E}\left[R_E^2\right]$, and $\rho={\mathop{\rm cov}} \left( {r_M^2, r_E^2} \right)/\sqrt{{\mathop{\rm var}} \left({r_M^2} \right){\mathop{\rm var}} \left( {r_E^2} \right)}$ is the power correlation coefficient of ($0\le \rho < 1$). $\rho$ is related to the correlation coefficient, $\rho_{G_M,G_E}$ of $G_M$ and $G_E$ by $\rho=\left|\rho_{G_M,G_E}\right|^2$. Now let the fading power gains $H_M$ and $H_E$ be defined by
\begin{eqnarray*}
  h_M=\xi_1\left(r_M, r_E\right)=\left|r_M\right|^2\,\, \textrm{and}\,\, h_E=\xi_2\left(r_M, r_E\right)=\left|r_E\right|^2,
\end{eqnarray*}
then the joint pdf of $H_M$ and $H_E$ can be obtained directly from the joint pdf of $R_M$ and $R_E$ using by the Jacobian of the transformation \cite{Leon-Garcia94Probability}:
\begin{eqnarray*}
  f_{H_M, H_E} \left( h_M, h_E \right) = f_{R_M, R_E} \left(\xi_1^{-1}\left(h_M, h_E\right), \xi_2^{-1}\left(h_M, h_E\right) \right)\left|\mathcal{J}\left(h_M, h_E\right)\right|,
\end{eqnarray*}
where $\left|\mathcal{J}\left(h_M, h_E\right)\right|$ is the Jacobian of the transformation defined by
\begin{eqnarray*}
\left|\mathcal{J}\left(h_M, h_E\right)\right| = \det \left[
{\begin{array}{*{20}c}
   {\partial \xi _1^{ - 1} /\partial h_M } & {\partial \xi _1^{ - 1} /\partial h_E }  \\
   {\partial \xi _2^{ - 1} /\partial h_M } & {\partial \xi _2^{ - 1} /\partial h_E }  \\
\end{array}} \right].
\end{eqnarray*}
Therefore the joint pdf of $H_M$ and $H_E$ is
\begin{align*}
  f_{H_M, H_E} \left( h_M, h_E \right) & =  f_{R_M, R_E} \left( \xi_1^{-1}\left(h_M, h_E\right), \xi_2^{-1}\left(h_M, h_E\right) \right)\left|\mathcal{J}\left(h_M, h_E\right)\right| \nonumber\\
                                       & =  \frac{{4 \sqrt{h_M h_E} }}{{\bar \gamma _M \bar \gamma _E \left({1 - \rho } \right)}}\exp \left[ { - \frac{{1 }}{{{1 - \rho }}}\left( {\frac{h_M}{{\bar \gamma _M }} + \frac{h_E}{{\bar \gamma _E}}} \right)} \right]I_0 \left( {\frac{{2 \sqrt{ \rho} \sqrt{h_Mh_E}}}{{ {1 - \rho }\sqrt{\bar \gamma _M \bar \gamma _E}}} } \right)\left|\frac{1}{4\sqrt{h_M h_E}}\right|\nonumber\\
                                       & = \frac{{1 }}{{\bar \gamma _M \bar \gamma _E \left( {1 - \rho } \right)}}\exp \left[ { - \frac{{1 }}{{{1 - \rho } }}\left( {\frac{h_M}{{\bar \gamma _M }} + \frac{h_E}{{\bar \gamma _E }}} \right)} \right]I_0 \left( {\frac{{2 }}{{ {1 - \rho }}}\sqrt{\frac{\rho h_M h_E}{{\bar \gamma _M \bar \gamma _E }}} } \right).
\end{align*}
%-------------------------------------------------------------------%
\section*{Acknowledgment}
The authors would like to thank Steven W. McLaughlin and Matthieu Bloch for their comments and suggestions, which improve this manuscript.
%-------------------------------------------------------------------%
\ifCLASSOPTIONcaptionsoff
  \newpage
\fi
%-------------------------------------------------------------------%
\bibliographystyle{IEEETran}
\bibliography{MyOwnBib}
%-------------------------------------------------------------------%
\end{document}